# Attaining scalable storage-expansion dualism for bioartificial tissues

S Sambu*

* Corresponding author:

P.O. Box 283

Nandi Hills, Kenya

Sambu@post.harvard.edu

## Abstract

The untenable dependence of cryopreservation on cytotoxic cryoprotectants has motivated the mining of biochemical libraries to identify molecular features marking cryoprotectants that may prevent ice crystal growth. It is hypothesized that such molecules may be useful across all temperatures eliminating cellular destruction due to equilibration cytotoxicity before and after cryopreservation. By probing the biochemical space using solvation-associated molecular topology and partition-distribution measures we developed an analytic formalism for ice crystal inhibition. By probing the union between a heat-shock protein (HSP) cluster and an anti-freeze glycoprotein (AFGP) cluster, the model development process generated distinct regions of interest (ROIs) for AFGPs and proved robust across different classes of proteins.  These results confirm that there is a chemical space within which efficacious Ice Crystal inhibitor (ICI) molecular libraries can be constructed. These spaces contain latent projections drawn from solvent accessibility, hydrogen bonding capacity and molecular geometry. They also showed that molecular design can be a useful tool in the development of new-age cryoactive matrices to enable the smooth transition across culture, preservation and expansion. These chemical spaces provide low cytotoxicity since such amphipathic molecules occupy a continuum between solubizing and membrane stabilizing regions as shown by the free energy of translocation calculations. These biochemical design spaces are fundamentally the solution to efficient scale-up and deployment of cell therapies to the patient's bedside. Consequently, this article proposes the use of a molecular knowledge-mining approach in the development of a class of non-cytotoxic cryoprotective agents in the class of ICI compatible with continuous cryothermic and normothermic cell storage and expansion.

## Introduction

Cryopreservation is concerned with the maintenance of biological cells, tissues, organs and constructs at very low temperatures sustained by cryogenic liquids and solids most often including liquid nitrogen, dry ice or even liquid helium. Prevailing theory attributes successful preservation to the suspended animation of biochemical reactions that sustain life via the reduction in the average kinetic energy of the biomolecules that underpin metabolism.

The slow cooling formalism states that slow cooling encourages ex-osmosis while CPA prevents ice crystallization; analogously, rapid warming encourages endosmosis and prevents ice recrystallization(1,2). This latter aspect is especially dependent on the construct size since large constructs will invariably fail to equilibrate forcing an upper size limit of about 600 µm (effective diffusion distance for small molecules < 100 Da)(3). It is the basic premise of this article that reducing ice formation with benign cryoprotectants designed according to knowledge-mining from biochemical spaces will push this limit much higher enabling the cryopreservation of larger constructs by reducing the time requirement for equilibration as well as the cytotoxicity during such processes.

ICI prevents colligative cellular and sub-cellular destruction. During re-warming cycles, as liquid water levels rise, ICI ensures the rapid attainment of an isotonic medium in advance of the restoration of metabolism. Therefore, ice crystal development under conditions reminiscent of heterogeneous nucleation is prevented, the typical persistence of high *colligative* stresses, deviations from target pH, solubility levels, ionic strengths and its effects on the denaturation of proteins and subcellular structures is inhibited(4).

It has also been theorized that ICI can prevent the development of shear stresses at the water-ice interface which would otherwise render further damage to proteins upon re-warming of constructs. The shear energies are positively correlated with increasing ice-water interfaces and can calculably be diminished by introducing amphipathic molecules with moieties to mediate between the ice-water regions and to minimise the interfacial tension(5).

The central role of ICI continues to be demonstrated in the field of cryopreservation. The primary argument is that successful cryopreservation and the development of benign non-advancing ice crystals below a determinable threshold is demonstrable and is known to be affected by factors such as cell size (6). In turn, cell size is linked to cell cycling (7) a known determinant of cellular osmotic parameters(8) which dictate the response events during cryopreservation. These interconnected factors confirm the strong link between culture and cryopreservation. These results have also demonstrated a definite tolerance of small, non-lethal ice crystals by cells which may indicate the role of mainstream cryoprotective agents in inhibiting Ostwaldian processes at critical stages of cryopreservation(1,9).

It has also been shown that certain polymers display ICI effects. These polymers can also be modified to deliver biomechanical signalling to cells in vitro to incept cell cycling among other cellular fate processes (3,10). Put together, such polymers can deliver a smooth transition between culture and cryopreservation.

In summary, ICI may enable a thermal and chemical continuum by providing a bulwark against *thermal and colligative stresses*. In so doing, ICI provide restorative biochemical processes during cryopreservation. Therefore, this article purposes to use a molecular knowledge-mining approach to develop a class of non-cytotoxic cryoprotective agents that will reduce equilibration and cryopreservation cytotoxicities enabling the development of protocols for symmetric and scalable storage-expansion for bioartificial tissues.

## Materials and Methods

### Data Collection:

Data was collected based on studies of ice crystal growth in various solutions of a class of chemistries known to inhibit recrystallization.
A second set of data was collected from UniProt (11) to help establish the range of classes of molecules that can inhibit the advancement of ice crystals. To contrast against ICI inhibitors, chaperone proteins/heat shock proteins (HSPs) that are endogenous to the hydrophilic cytosol but must necessarily possess some hydrophobic sub-domains were

included(12). To date, HSPs are not directly involved in inhibiting ice crystal formation but are evolutionarily involved in response to osmotic shock and dehydration events by interacting with endangered intracellular proteins at risk of unchecked protein unfolding (13). Libraries for both classes were drawn from the RSCB Protein Data Bank (14).

### Generation of molecular descriptors

Molecular descriptors were generated using molecular graph theory; supramolecular descriptors were drawn from semi-empirical models; finally, models were developed in R using the cheminformatics packages.
The RCDK & RCPI (R/Bioconductor) packages(15) were used to obtain chemical descriptors including hydrogen bond donors and acceptors, partitioning and distribution coefficients. Surface area calculations were based on the Volume Area Dihedral Angle Reporter(16) (VADAR); upstream of the VADAR simulation, the PDB quality was assessed using ResProx(17).

### Computational techniques

Initially, a logistic regression approach was used to test the base hypothesis for the key factors driving HSP and AGFP behaviours. Factors that are known to be critical in selecting ICI molecules were thereafter used unless the data showed minimal sensitivity. The R-GLM Package was used for this work. Relationships were mined using the R-PLS package. Finally, for the direct computation of free energy of translocation for AFGP/HSP molecules, EMBOSS (European Molecular Biology Open Software Suite)(18) was deployed; hypothesis testing for differences was run in R

### Results and Discussion

A chemical series was mined to coincide with the ICI molecular space and with a growth in complexity and in the ICI response. The ICI response is often measured by using the mean grain size of the advancing ice crystal. An RSCB set of 562 molecules were used to train the model. To initiate the analysis, a logistic regression is run to determine which factors drive a sharp classification boundary between the two classes; results on Figure 1 show that solvation and bonding descriptors are preponderant.

### Bonding descriptors

It is worth noting that the usefulness of geometry, topology and charge distribution (such as solvent accessible surface areas and polar surface areas) in determining the success in ICI hydrogen bonding or van der Waals forces increases when bonding descriptors are included in the model. The significance of hydrogen bonding is supported by the results of the logistic regression shown in Figure 1 – represented as hydrogen bond donors and acceptor counts. This is consistent with the findings in the solvation descriptors. Subsequently, **Error! Reference source not found.** shows a variable importance plot of the AFGP and HSP data sets with the HBDA (summation of Hydrogen Bond Donors and Acceptors, see Equation 1) coming in either second or third place overall. HBDA as a measure of hydrogen bonding capacity is derived using Equation 1. In Figure 1 and Figure 2, the techniques demonstrate the importance of bonding descriptors in an aqueous environment.

Equation 1: Hydrogen Bonding Capacity (HBC) Equation

$$HBDA = \sum_{i=1}^{n} HBA_i + HBD_i$$

```
Coefficients:
                    Estimate Std. Error z value Pr(>|z|)
(Intercept)       -26.142343   4.393482  -5.950 2.68e-09 ***
SAscore             0.159147   0.094630   1.682   0.0926 .
SAscore_Fragments  44.629100   8.285977   5.386 7.20e-08 ***
ALogP              -0.048880   0.010433  -4.685 2.80e-06 ***
LogD               -0.028516   0.004933  -5.781 7.44e-09 ***
HBDA               -0.008010   0.001543  -5.191 2.09e-07 ***
---
Signif. codes:  0 '***' 0.001 '**' 0.01 '*' 0.05 '.' 0.1 ' ' 1
```

Figure 1: A Results from a logistic regression for HSP and AFGP response classes showing the importance of solvation and hydrogen bonding (alternatively non-covalent hydrophilic-hydrophobic interactions).

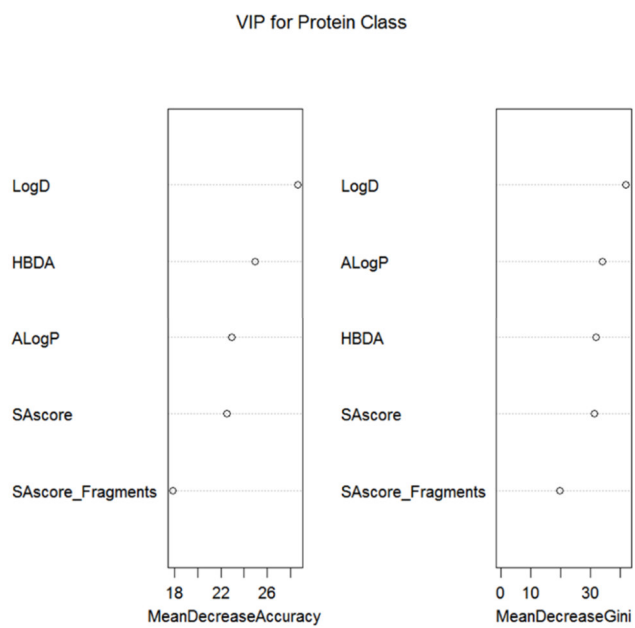

Figure 2: A variable importance plot showing the importance of solvation and bonding parameters for the AFGP and HSP data sets.

## Solvation and Partitioning descriptors

It would have been surprising not to have a measure of solvation in the descriptor properties since the interaction of the ICI agents with water in both solid and liquid phases is essential. This is confirmed by the significance of parameters for solvent accessibility at a fragmentary and molecular level as shown by Figure 3 whereby AFGPs demonstrate a greater solvent accessibility and such feature can be expected to translate to molecular attributes of an intermediary compound. As a counterpoint, the distribution coefficient of HSPs will tend to favour an aqueous phase (negative values) while the AFGPs will have a tendency towards the neutral null position reflecting amphipathic features as shown in Figure 4A.
To improve separability, invoking the molecular geometry using the allometric measure normalizing molecular area to molecular volume (SAV) demonstrably improves separation in Figure 4B. there is still a zone of overlap showing that some HSP can have molecular descriptors like AFGPs. In fact, it has been empirically shown that HSPs can help to protect intracellular environments from ice crystal growth (19). On the latter piece, the fragmentary attributes confirm the need to distinguish between polar and non-polar faces of the molecule for AFGPs/ICI molecules while for HSPs, their chaperoning activity requires polar faces mediating between proteins and solvent milieus – this feature is demonstrable in Figure 4. In Figure 5, it can be shown that HSPs have more hydrogen bonding interactions but with a lower (more negative) solvent accessibility score than AFGPs. This corresponds with the amphipathy of AFGPs and the tendency for AFGPs to expose constituent fragments to the surrounding environment. To confirm these trends, a comparison of the free energy of translocation of HSP and AFGP molecules was run in EMBOSS; Figure 6 shows that the HSP molecules have a lower translocation energy than AFGPs which can be attributed to the amphipathy of AFGPs. These observations are consistent with those seen in nature for AFGPs (20) and the observations of buried hydrophobic pocket residues and relatively exposed hydrophilic residues for HSPs (21). All together, these results show that there can be a synergy attainable by using HSPs and AFGPs in concert whereby cytosolic protections are afforded by HSPs while membrane protections are secured using AFGPs.

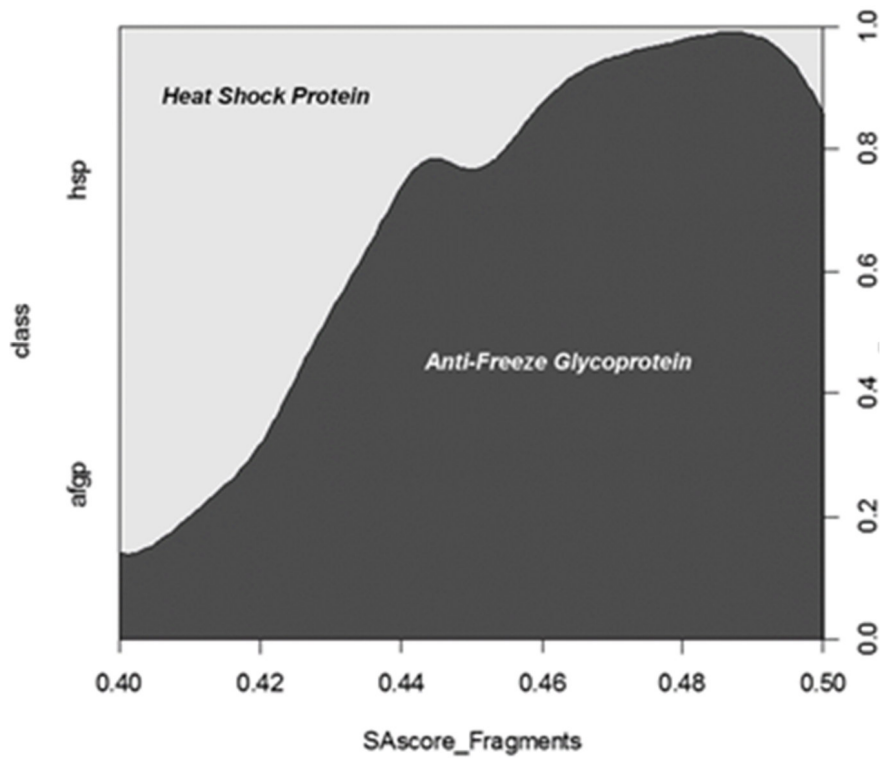

Figure 3: Solvent accessible Score for Fragments (SASF, "SAscore_Fragments") axis shows that a high solvent accessibility score drives assignment to the AFGP class. This is consistent with current theory that the AFGP class bonds to a hydrophilic ice surface while exposing a hydrophobic facet into the liquid phase preventing further ice crystal advancement; in contrast, HSP are chaperoning molecules requiring constant intermediation between the protein and aqueous cytosol. Overall, a 0.43 boundary appears to divide the conditional distribution into HSP-rich and AFGP-rich zones.

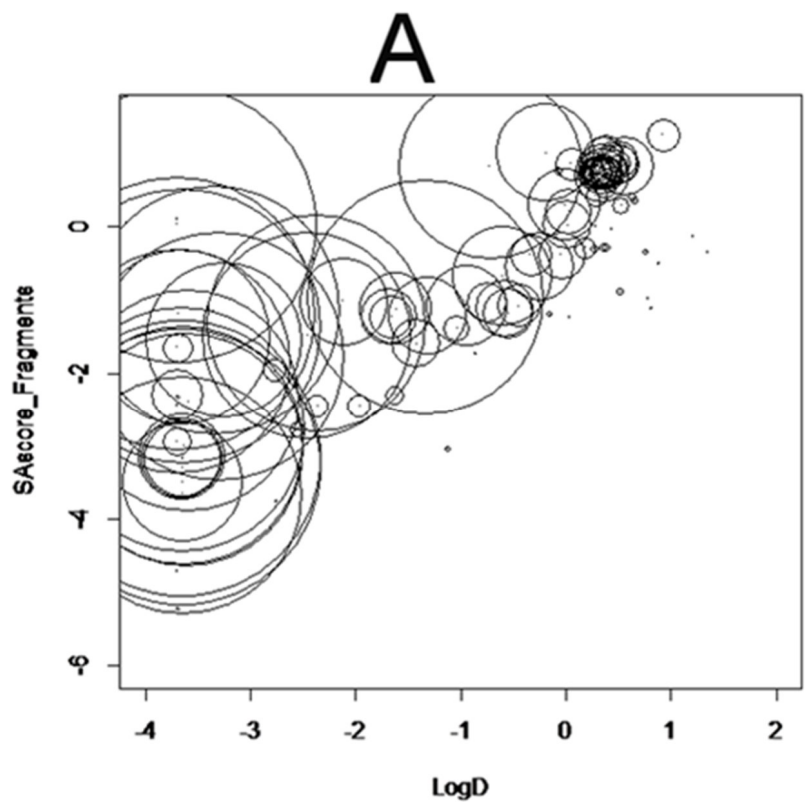
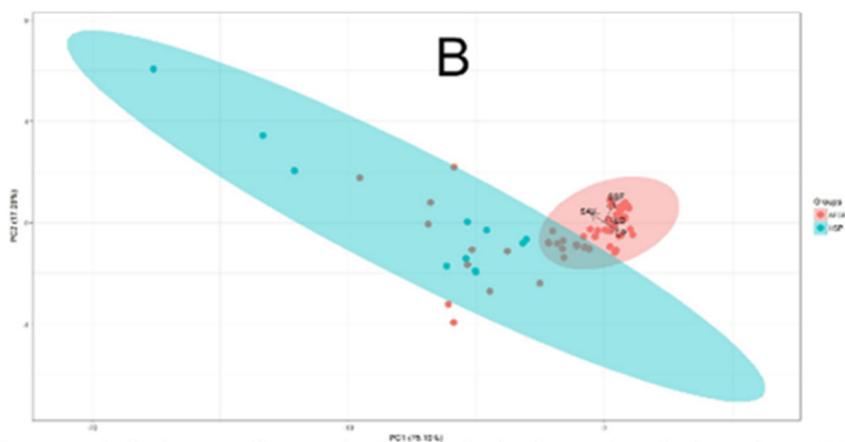

Figure 4: (A) Evaluation of the impact of a distribution coefficient on the probability for classification of a molecule as either an AFGP or an HSP using a nominal logistic regression (NLR). The circles represent probabilities for a molecule belonging to an AFGP class. Large circles overlap the smaller circles and may imply the two factors alone do not allow separability between the classes; it may therefore point to a molecular descriptor overlap between the two classes (visualized in "B"). this overlap may indicate the ability to design a chaperone functionality (typical of HSP) as well as an ice crystal inhibitor as is typical of AFGPs from within the union of AFGPs and HSPs

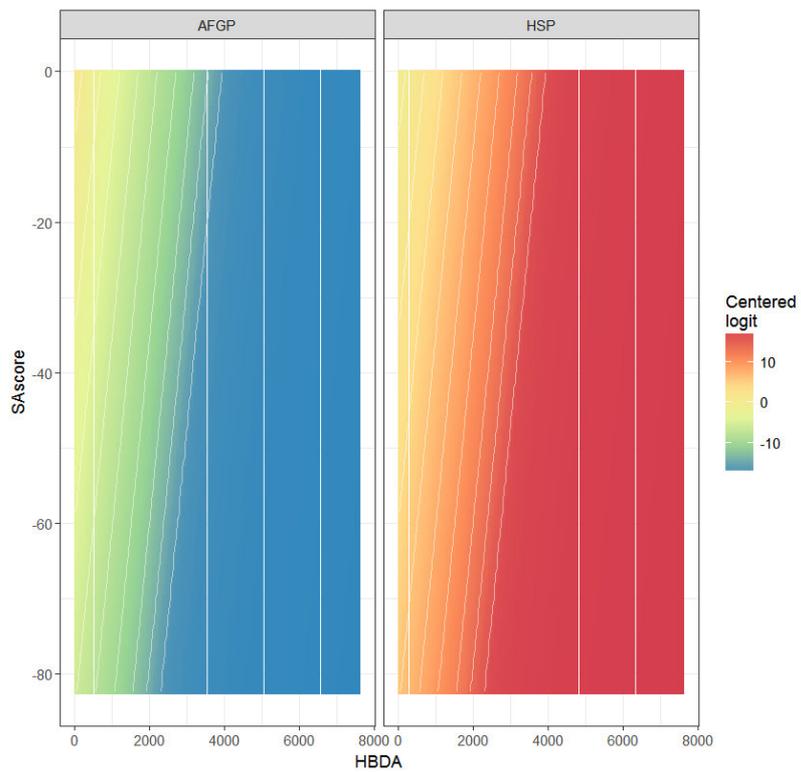

Figure 5: HSPs have more hydrogen bonding interactions but with a lower (more negative) solvent accessibility score than AFGPs. This corresponds with the amphiphilicity of AFGPs and the tendency for AFGPs to expose constituent fragments to the surrounding environment

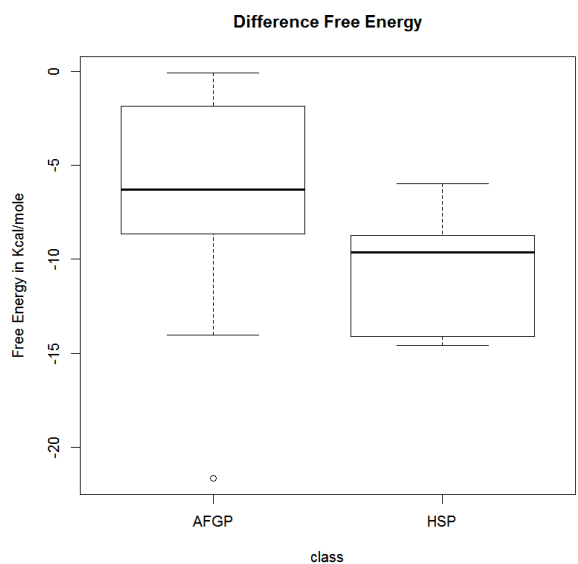

Figure 6: A free energy calculation using EMBOSS showing the favoured partitioning of HSP into a lipid core is, on average, greater than that of AFGP. A Welch 2-sample t-test showed there is a significant difference in the free energy of transition (p-value < 2.2e-16)

## Conclusion

In conclusion, a molecular-theoretic approach to predicting cryopreservation outcomes may lend itself well in the design of next generation cryogenic media with a lower cytotoxicity and higher yields. AFGPs are shown to be solubizing or stabilizing compounds with relatively more hydrophobic attributes relative heat shock proteins. Both being rather benign molecules with cryoprotective and culture-proofing benefits respectively can be employed in an expansion and preservation process that can scale up production and deliver steady supplies of cells and artificial constructs for cell therapy. It is proposed that future work may focus on the mathematical treatment of cell-biomaterial boundaries in a gradual push from a strong chemical and material science foundation to a more composite, biomedical theory for advancing the development of a continuous supply mechanism for cell therapy.